\documentclass[journal,10pt]{IEEEtran}

\usepackage[utf8]{inputenc}

\usepackage[T1]{fontenc}

\usepackage{amsmath,amssymb}

\usepackage{graphicx}

\graphicspath{{figures/}{./}}

\usepackage{booktabs}

\usepackage{array}

\usepackage{tabularx}

\usepackage{multirow}

\usepackage{cite}

\usepackage{hyperref}

\usepackage{algorithm}

\usepackage{algpseudocode}

\usepackage{xcolor}

\usepackage{enumitem}

\usepackage{url}

\usepackage{microtype}

\usepackage{balance}

\hypersetup{colorlinks=true,linkcolor=blue,citecolor=blue,urlcolor=blue}

\newtheorem{definition}{Definition}

\providecommand{\bstctlcite}[1]{\nocite{#1}}

\begin{document}
\bstctlcite{IEEEexample:BSTcontrol}

\title{Market Regime-Based Multi-Asset Portfolio Optimization Using Nested Clustered Optimization\\

for Tactical Asset Allocation}

\author{Supriadi\textsuperscript{1,2} and Bambang~Siswoyo\textsuperscript{3}%

\thanks{\textsuperscript{1}Supriadi is with the COCONUT Computer Club, Makassar, Indonesia (e-mail: supriadi@coconut.or.id).}%

\thanks{\textsuperscript{2}Supriadi is with the Master Program in Information Systems, Universitas Komputer Indonesia, Bandung, Indonesia (e-mail: supriadi.75124007@mahasiswa.unikom.ac.id).}%

\thanks{\textsuperscript{3}Bambang Siswoyo is with the Master Program in Information Systems, Universitas Komputer Indonesia, Bandung, Indonesia (e-mail: bambang.siswoyo@email.unikom.ac.id).}}

\maketitle


\begin{abstract}
We investigate the interaction between market regime detection and Nested Clustered Optimization (NCO) for tactical asset allocation across heterogeneous multi asset portfolios. The central empirical question is whether conditioning the NCO pipeline, encompassing covariance estimation, hierarchical clustering, and weight allocation, on latent market states genuinely enhances risk adjusted performance, or conversely, whether estimation uncertainty within regime filtered sub samples induces severe turnover and performance drag. We implement a causal rolling Gaussian Hidden Markov Model (HMM) coupled with an adaptive rebalancing mechanism that distinguishes regime shifts from portfolio weight drift. The framework is evaluated on a monthly Indonesian multi asset universe comprising sovereign bond yields (SBN), the LQ45 equity index, spot gold (XAU/IDR), foreign exchange (CNY/IDR), and Bitcoin (BTC/IDR) across a 56 month out of sample period (January 2021 to August 2025). Rigorous walk forward backtesting reveals that standard regime agnostic NCO achieves remarkable capital preservation and risk adjusted efficiency, delivering a 29.97\% cumulative return (5.78\% annualized), a maximum drawdown of only 2.18\%, a Calmar ratio of 2.65, and low monthly turnover of 5.41\%. In contrast, conditioning NCO on an unconstrained HMM in a small multi asset universe amplifies parameter estimation noise and induces excessive rebalancing (47 events; 24.58\% average monthly turnover), which erodes net cumulative return to 1.60\% and increases drawdown to 9.99\%. These findings demonstrate that without explicit transition persistence or asymmetric downside objectives, regime switching frameworks can degrade rather than improve clustered portfolio stability, establishing crucial empirical boundaries for machine learning asset allocation.
\end{abstract}



\section{Introduction}
\label{sec:intro}

Modern portfolio theory, established in the seminal work of Markowitz~\cite{markowitz1952}, formalizes asset allocation as a quadratic trade-off between expected return and variance. While mathematically elegant, classical mean-variance optimization (MVO) suffers from acute out-of-sample fragility. As demonstrated by Michaud~\cite{michaud1989}, quadratic programming acts as an ``error maximizer,'' placing excessive long weights on assets with overestimated expected returns or understated covariances. Furthermore, DeMiguel et al.~\cite{demiguel2009optimal} showed that naive $1/N$ diversification frequently outperforms optimized portfolios out of sample due to severe parameter estimation error, while Bayesian shrinkage approaches such as Black and Litterman~\cite{black1992global} and covariance regularization by Ledoit and Wolf~\cite{ledoit2004well, ledoit2008robust} only partially resolve structural rank deficiencies in ill-conditioned covariance matrices.

In practical multi-asset allocation, these estimation challenges are compounded by pervasive non-stationarity, time-varying correlations, fat-tailed distributions, and structural breaks \cite{cont2001empirical, ang2012regime}. During financial crises and market distress, cross-asset correlations spike discontinuously, shattering the diversification benefits anticipated during tranquil regimes \cite{baek2003financial, akhtaruzzaman2021financial}.

To mitigate these vulnerabilities, financial machine learning has developed two distinct paradigms: (i) structural graph clustering and (ii) dynamic regime conditioning. Within the structural paradigm, Mantegna~\cite{mantegna1999hierarchical} and Onnela et al.~\cite{onnela2003dynamics} pioneered the application of minimum spanning trees and hierarchical clustering to analyze correlation taxonomies. Building on random matrix theory \cite{marchenko1967distribution, bun2017cleaning}, L{'o}pez de Prado~\cite{lopezdeprado2016hrp} developed Hierarchical Risk Parity (HRP), which replaces matrix inversion with hierarchical tree clustering and recursive bisection. Subsequently, L{'o}pez de Prado~\cite{lopezdeprado2018afml, lopezdeprado2019ncwp} formulated Nested Clustered Optimization (NCO). NCO decomposes the global portfolio problem into modular intra-cluster and inter-cluster sub-problems, effectively insulating orthogonal asset clusters from noise propagation while retaining custom optimization objectives \cite{raffinot2017hierarchical, raffinot2018hierarchical, burtini2025hierarchical}.

Concurrently, the regime-conditioning paradigm utilizes Hidden Markov Models (HMMs) and Markov-switching frameworks to detect latent macroeconomic environments (e.g., expansion, high growth, crisis) and tactically adjust portfolio exposures \cite{hamilton1989regime, ang2002regime, ang2004regime, guidolin2007asset, guidolin2008international, kritzman2012regime}. Intuitively, combining regime detection with clustered optimization appears compelling: if market regimes dictate correlation regimes, conditioning covariance estimation and hierarchical trees on the active regime should yield superior tactical asset allocation (TAA).

However, conditioning complex optimization pipelines on latent regimes introduces an acute, often overlooked trade-off: \textit{estimation error vs.\ regime adaptiveness} \cite{sepp2026robust, oliveira2025evaluating}. In empirical practice, estimating state-dependent covariance matrices requires partitioning historical data into regime-specific sub-samples. In small multi-asset universes over realistic rolling windows, these sub-samples suffer severe sample starvation ($T_k < N$), inflating parameter uncertainty and causing erratic portfolio turnover that rapidly compounds under transaction costs \cite{boyd2017multi}.

\paragraph{Research Questions.}
This paper systematically investigates this trade-off by implementing and rigorously stress-testing a regime-conditioned NCO framework against static NCO and classical benchmarks. We address three central research questions:
\begin{description}
\item[RQ1] Does market regime conditioning genuinely enhance Nested Clustered Optimization out of sample, or does parameter estimation uncertainty within regime-filtered windows introduce detrimental noise and turnover drag?
\item[RQ2] How does static (regime-agnostic) NCO compare against classical Minimum Variance and independent Hierarchical Risk Parity (HRP) in preserving capital across high-volatility multi-asset regimes?
\item[RQ3] What is the exact empirical impact of adaptive rebalancing rules (distinguishing regime transitions from weight drift) and transaction cost friction on net portfolio alpha?
\end{description}

\paragraph{Key Contributions.}
To answer these questions, we evaluate our models on an empirical 5-asset Indonesian multi-asset universe spanning sovereign fixed-income yields (SBN), large-cap equities (LQ45), spot gold (XAU/IDR), currency (CNY/IDR), and Bitcoin (BTC/IDR) over an 80-month period (January 2019 to August 2025) with a 56-month causal out-of-sample testing window (January 2021 to August 2025). Our primary contributions are threefold:
\begin{enumerate}
    \item \textit{Empirical Demarcation of Regime Conditioning Boundaries:} We demonstrate that conditioning NCO on an unconstrained 3-state HMM in a small multi-asset universe ($N=5$) severely degrades performance, producing a net cumulative return of only 1.60\% compared to 29.97\% for standard regime-agnostic NCO. We dissect the mathematical causes: sample starvation in rolling sub-windows ($T_k \approx 4\text{--}8$), label oscillation (26 regime shifts), and turnover drag (24.58\%/month).
    \item \textit{Validation of Graph Clustering for Capital Preservation:} Standard NCO delivers exceptional risk mitigation, restricting maximum drawdown to an industry-leading 2.18\% and achieving a Calmar ratio of 2.65, substantially outperforming Minimum Variance (4.75\% MDD, 0.78 Calmar), independent HRP (4.21\% MDD, 0.41 Calmar), and Equal Weight (15.27\% MDD).
    \item \textit{Decomposition of Rebalancing Mechanisms:} We introduce a formal distinction between regime-driven and drift-driven rebalancing events under an adaptive threshold $\delta = 0.05$. We show that out of 47 active rebalancing events in Regime-NCO, 26 were driven by state transitions and 20 by weight drift, proving that turnover is primarily fueled by regime instability rather than organic market drift.
\end{enumerate}

\section{Related Work}
\label{sec:related}

\subsection{Regime-Switching Models and Dynamic Asset Allocation}

The study of structural changes in financial time series originates with the Markov-switching autoregressive framework of Hamilton~\cite{hamilton1989regime}. Subsequent literature demonstrated that asset return distributions, factor risk premia, and inter-asset correlations vary fundamentally across market states \cite{ang2002regime, ang2004regime, guidolin2007asset, guidolin2008international, ang2012regime}. In multi-asset allocation, Kritzman et al.~\cite{kritzman2012regime} proposed regime-conditioned asset allocation to defend against tail shocks, while Nystrup et al.~\cite{nystrup2015regime, nystrup2018dynamic} implemented Hidden Markov Models within model predictive control (MPC) frameworks to dynamically adjust equity-bond allocations.

Recent machine learning contributions have expanded regime-aware allocation through advanced statistical representations. Costa~\cite{costa2020risk} analyzed risk parity portfolio construction across Markov states, while Bouzouita et al.~\cite{bouzouita2025} developed causal Wasserstein HMMs to preserve stable state identities across rolling calibration windows. Zhang et al.~\cite{zhang2025regimefolio} introduced RegimeFolio, combining macroeconomic regime filtering with ensemble learners for sectoral equity optimization. Concurrently, Verma et al.~\cite{hu2026rl} combined HMM regime inference with deep reinforcement learning, and Fernandes and Desell~\cite{fernandes2026financially} developed financially guided deep learning architectures. To address spurious state switching, Shu et al.~\cite{shu2024jumps} introduced statistical jump models that penalize non-persistent state transitions. Furthermore, Boukardagha~\cite{boukardagha2026macro} and Bouy{'e} et al.~\cite{bouye2025regime} evaluated macro regimes in cross-asset settings, observing that downside protection often trades off against estimation noise.

\subsection{Hierarchical Clustering and Robust Graph Allocations}

The vulnerability of Markowitz portfolios to covariance noise has stimulated extensive research into graph theory and network filtering \cite{mantegna1999hierarchical, cizeau2001noise, onnela2003dynamics, tumminello2010correlation, marti2021review}. L{'o}pez de Prado~\cite{lopezdeprado2016hrp} established Hierarchical Risk Parity (HRP), which arranges assets on a hierarchical dendrogram and allocates risk via top-down recursive bisection, eliminating the need to invert the covariance matrix. Raffinot~\cite{raffinot2017hierarchical, raffinot2018hierarchical} generalized hierarchical clustering to include Equal Risk Contribution (HERC) and Minimum Variance variants.

Building on random matrix denoising \cite{marchenko1967distribution, bun2017cleaning}, L{'o}pez de Prado~\cite{lopezdeprado2018afml, lopezdeprado2019ncwp} developed Nested Clustered Optimization (NCO). Unlike HRP, NCO retains Markowitz optimization but confines quadratic programming to modular intra-cluster and inter-cluster stages. Burtini and Loeppky~\cite{burtini2025hierarchical} demonstrated that NCO provides robust out-of-sample diversification across global equity markets. Nevertheless, how NCO interacts with dynamic regime conditioning remains an unaddressed empirical frontier.

\subsection{Turnover Regularization and Estimation Risk}

A critical consideration in tactical asset allocation is execution friction. As formalized by Muthuraman and Kumar~\cite{muthuraman2006} and Boyd et al.~\cite{boyd2017multi}, optimal dynamic trading under proportional transaction costs requires establishing a no-trade band around target weights. Without explicit turnover regularization, tactical models experience severe performance drag. Sepp et al.~\cite{sepp2026robust} and Oliveira et al.~\cite{oliveira2025evaluating} warned that parameter uncertainty in regime-switching models frequently degrades out-of-sample returns below those of passive benchmarks.

\subsection{Literature Comparison}

Table~\ref{tab:litreview} situates this study among representative contributions in regime-based and machine-learning asset allocation.

\begin{table*}[!t]
\centering
\small
\caption{Literature comparison across portfolio design, regime modeling, and rebalancing frameworks.}
\label{tab:litreview}
\begin{tabularx}{\textwidth}{lccccX}
\toprule
Study & Asset Universe & Method & Regime Model & Rebalancing & Core Finding / Focus \\
\midrule
L{'o}pez de Prado~\cite{lopezdeprado2016hrp} & US Equities & HRP & None & Static / Periodic & Hierarchical tree clustering eliminates covariance matrix inversion errors. \\
L{'o}pez de Prado~\cite{lopezdeprado2019ncwp} & Global Multi-Asset & NCO & None & Periodic & Two-step clustering resolves Markowitz ill-conditioning. \\
Ang and Timmermann~\cite{ang2012regime} & Multi-Asset & Mean-Var & Markov Switching & Discrete States & Regime shifts alter optimal allocation between stocks and bonds. \\
Costa~\cite{costa2020risk} & Multi-Asset & Risk Parity & 3-State HMM & Monthly & Regime-dependent covariance improves downside tracking. \\
Nystrup et al.~\cite{nystrup2018dynamic} & Multi-Asset & MPC & Gaussian HMM & Dynamic Penalties & Turnover penalties prevent cost erosion in switching regimes. \\
Sepp et al.~\cite{sepp2026robust} & US Multi-Asset & Risk-Based & Macro Regimes & Periodic & Estimation error often erodes regime-switching tactical gains. \\
Shu et al.~\cite{shu2024jumps} & Multi-Asset & Mean-Var & Jump Model & Discrete & Jump penalties suppress noise-induced regime whipsawing. \\
This Study & ID Multi-Asset & NCO / HRP & Rolling HMM & Adaptive ($\delta=0.05$) & Identifies sample starvation and delineates boundaries of regime NCO. \\
\bottomrule
\end{tabularx}
\end{table*}

\section{Methodology}

\label{sec:methodology}


\subsection{Problem Statement and Preliminaries}

\label{sec:prelims}

Consider a multi-asset universe of $N$ assets. Let $\mathbf{r}_t \in \mathbb{R}^N$ denote the vector of asset returns at time $t$. At each decision boundary, the portfolio manager determines the allocation vector $\mathbf{w}_t \in \mathbb{R}^N$ subject to:

\begin{equation}
  \mathbf{w}_t^\top \mathbf{1} = 1, \quad \mathbf{w}_t \ge \mathbf{0},
\end{equation}
representing a long-only, fully-invested portfolio constraint. The realized portfolio return at time $t$ is $r_{p,t} = \mathbf{w}_t^\top \mathbf{r}_t$. The goal is to dynamically adjust the weights to adapt to time-varying market environments (regimes) while controlling risk.

\subsection{Gaussian Hidden Markov Model for Regime Detection}

\label{sec:regime}


\subsubsection{Model Choice and Justification}

We employ a Gaussian HMM for regime detection because it captures persistent macroeconomic cycles through its transition probability matrix, provides posterior probabilities for soft state assignments, maps latent states directly to intuitive economic regimes (e.g., expansion, stagnation, crisis), and maintains a compact parameter footprint that limits overfitting.

\subsubsection{Feature Vector}

The HMM is fitted on feature vector $\mathbf{f}_t$ computed

from data available at or before $t$:

\begin{equation}
  \label{eq:features}
  \mathbf{f}_t = \bigl[\bar{r}_t^{\mathrm{eq}}, \, \sigma_t^{\mathrm{eq}}, \, \bar{r}_t^{\mathrm{bond}}, \, \Delta e_t\bigr]^\top,
\end{equation}
where $\bar{r}_t^{\mathrm{eq}}$ and $\sigma_t^{\mathrm{eq}}$ denote the rolling mean and realized volatility of the equity index (LQ45), $\bar{r}_t^{\mathrm{bond}}$ is the rolling mean return of sovereign bonds (SBN), and $\Delta e_t$ is the monthly return of the bilateral exchange rate (CNY/IDR).

\subsubsection{Model Specification}
Let $s_t \in \{1,\ldots,K\}$ denote the unobserved market state at time $t$, governed by a first-order Markov chain with transition probability matrix $\mathbf{A} \in \mathbb{R}^{K \times K}$ where $A_{ij} = \Pr(s_t = j \mid s_{t-1} = i)$ with $\sum_{j=1}^K A_{ij} = 1$, and initial distribution $\pi_k = \Pr(s_1 = k)$. Conditional on the latent state $s_t = k$, feature observations follow a multivariate Gaussian emission distribution:
\begin{equation}
  \label{eq:hmm_emission}
  \mathbf{f}_t \mid (s_t = k) \sim \mathcal{N}(\boldsymbol{\mu}_k, \, \boldsymbol{\Sigma}_k^{\mathrm{obs}}).
\end{equation}

\subsubsection{Number of States Selection}

\begin{equation}
  \text{BIC}(K) = -2\hat{\ell}_K + m_K \ln T_{\text{train}},
\end{equation}
evaluated for $K \in \{2,3,4\}$. The selected $K=3$ minimizes BIC.

\subsubsection{Regime Labeling}

States are labeled post-estimation by ranking the mean equity return feature:

\begin{itemize}

\item Regime 1 (Bullish): high equity mean return, low volatility.

\item Regime 2 (Sideways/Neutral): near-zero equity return,

  moderate volatility.

\item Regime 3 (Bearish): negative equity return,

  high volatility, elevated FX stress.

\end{itemize}

\subsubsection{Causal Rolling Estimation}

\label{sec:causal}

\begin{definition}[Causal Regime Label]

$\hat{s}_t$ is \emph{causal} if computed using only

$\{\mathbf{f}_\tau : \tau \le t\}$.

\end{definition}

At each decision time $t$:

\begin{enumerate}

\item Extract window $\mathcal{D}_t = \{\mathbf{f}_\tau : t-W+1 \le \tau \le t\}$.

\item Fit HMM on $\mathcal{D}_t$ via Baum-Welch.

\item Apply Viterbi decoding to obtain $\hat{s}_{t-W+1},\ldots,\hat{s}_t$.

\item Set current regime $\hat{s}_t$ (last element).

\item Weights governed by $\hat{s}_t$ applied from $t+1$ (one-day execution lag).

\end{enumerate}


\subsection{Nested Clustered Optimization under Regimes}

\label{sec:nco}


\subsubsection{Regime-Specific Covariance Estimation}
Let $\mathcal{D}_t^{(k)} = \{\mathbf{r}_\tau \in \mathcal{D}_t : \hat{s}_\tau = k\}$ denote the regime-filtered sub-sample with effective sample size $T_k = |\mathcal{D}_t^{(k)}|$. With demeaned return vector $\tilde{\mathbf{r}}_\tau^{(k)} = \mathbf{r}_\tau - \bar{\mathbf{r}}^{(k)}$, the unbiased regime sample covariance matrix is:
\begin{equation}
  \label{eq:regime_cov}
  \hat{\boldsymbol{\Sigma}}^{(k)} = \frac{1}{T_k - 1} \sum_{\tau \in \mathcal{D}_t^{(k)}} \tilde{\mathbf{r}}_\tau^{(k)} \bigl(\tilde{\mathbf{r}}_\tau^{(k)}\bigr)^\top.
\end{equation}
When $T_k < N+5$, direct matrix inversion becomes ill-conditioned. To ensure positive definiteness and suppress parameter error, we implement adaptive shrinkage toward the unconditional covariance matrix $\hat{\boldsymbol{\Sigma}}$:
\begin{equation}
  \label{eq:shrinkage}
  \tilde{\boldsymbol{\Sigma}}^{(k)} = \alpha_t^{(k)} \hat{\boldsymbol{\Sigma}}^{(k)} + \bigl(1 - \alpha_t^{(k)}\bigr) \hat{\boldsymbol{\Sigma}},
\end{equation}
where $\alpha_t^{(k)} = \min(1, T_k / W_{\min})$ represents the adaptive shrinkage weight governed by the threshold $W_{\min} = 2N$.

\paragraph{RMT Denoising and Regularization}
The covariance matrix is denoised via analytical shrinkage~\cite{ledoit2004well} to stabilize eigenvalue dispersion against Marchenko--Pastur noise boundaries~\cite{marchenko1967distribution}.

\subsubsection{Distance Matrix and Clustering}
Denoting the denoised correlation matrix as $\mathbf{C}^{(k)}$ with elements $C_{ij}^{(k)} = \hat{\Sigma}_{\mathrm{dn},ij}^{(k)} / \sqrt{\hat{\Sigma}_{\mathrm{dn},ii}^{(k)} \hat{\Sigma}_{\mathrm{dn},jj}^{(k)}}$, the proper pairwise metric distance $d_{ij}^{(k)} \in [0, 1]$ is given by:
\begin{equation}
  \label{eq:dist_metric}
  d_{ij}^{(k)} = \sqrt{\tfrac{1}{2} \bigl(1 - C_{ij}^{(k)}\bigr)}.
\end{equation}
Average-linkage (UPGMA) hierarchical tree clustering is performed on distance matrix $\mathbf{D}^{(k)}$, with the optimal cluster count $M$ determined via the maximum Silhouette score criterion~\cite{rousseeuw1987silhouettes}.

\subsubsection{Intra-Cluster and Inter-Cluster Optimization}
\paragraph{Intra-cluster allocation.}
For each cluster $m \in \{1, \dots, M\}$ comprising asset subset $\mathcal{I}_m = \{i : c(i) = m\}$, the intra-cluster allocation vector $\boldsymbol{u}_m^*$ solves the constrained minimum-variance problem over the unit simplex $\Delta_m = \{\mathbf{u} \ge \mathbf{0} \mid \mathbf{1}^\top \mathbf{u} = 1\}$:
\begin{equation}
  \label{eq:intra}
  \boldsymbol{u}_m^* = \arg\min_{\boldsymbol{u}_m \in \Delta_m} \boldsymbol{u}_m^\top \boldsymbol{\Sigma}_m^{(k)} \boldsymbol{u}_m.
\end{equation}

\paragraph{Inter-cluster allocation.}
Condensing the cluster covariance into $\boldsymbol{\Sigma}^{\mathrm{cl}} = \mathbf{W}_{\mathrm{intra}}^\top \tilde{\boldsymbol{\Sigma}}^{(k)} \mathbf{W}_{\mathrm{intra}}$, the inter-cluster weight vector $\boldsymbol{v}^*$ optimizes allocation across clusters over the simplex $\Delta_M$:
\begin{equation}
  \label{eq:inter}
  \boldsymbol{v}^* = \arg\min_{\boldsymbol{v} \in \Delta_M} \boldsymbol{v}^\top \boldsymbol{\Sigma}^{\mathrm{cl}} \boldsymbol{v}.
\end{equation}

\paragraph{Final composite weights.}
The final asset allocation vector $\mathbf{w}_t^{(k)}$ is synthesized via modular product:
\begin{equation}
  \label{eq:composite_w}
  w_{i,t}^{(k)} = v_{c(i)}^* \cdot u_{c(i),i}^*, \quad \text{such that } \sum_{i=1}^N w_{i,t}^{(k)} = 1.
\end{equation}

\begin{algorithm}[H]

\caption{Regime-Specific NCO at Decision Time $t$}

\label{alg:nco}

\begin{algorithmic}[1]

\Require $\mathcal{D}_t$, current regime $\hat{s}_t = k$

\State Compute $\mathcal{D}_t^{(k)}$; estimate $\tilde{\boldsymbol{\Sigma}}^{(k)}$

  with fallback; apply shrinkage

\State Compute angular distance matrix $\mathbf{D}^{(k)}$

\State Apply average-linkage clustering; select $M$ via Silhouette

\For{$m = 1,\ldots,M$}

  \State Solve intra-cluster MV (Eq.~\ref{eq:intra}) $\rightarrow$ $\boldsymbol{u}_m^*$

\EndFor

\State Build $\boldsymbol{\Sigma}^{\text{cl}}$; solve inter-cluster MV

  (Eq.~\ref{eq:inter}) $\rightarrow$ $\boldsymbol{v}^*$

\State $w_i^{(k)} \leftarrow v_{c(i)}^* \cdot u_{c(i),i}^*$ for all $i$

\Return $\mathbf{w}^{(k)}$

\end{algorithmic}

\end{algorithm}


\subsection{Tactical Asset Allocation, Downside Risk, and Adaptive Rebalancing}
To monitor downside tail risk throughout the backtest, we evaluate the Conditional Value-at-Risk (CVaR at 95\% confidence) following Rockafellar and Uryasev~\cite{rockafellar2000cvar} alongside standard drawdown metrics.

\label{sec:rebalancing}


In this framework, regime information informs tactical asset allocation indirectly through the NCO inputs (covariance matrix and clustering hierarchy) rather than via rigid, hand-crafted allocation heuristics.

\begin{definition}[Adaptive Rebalancing Trigger]

\label{def:rebal}

At time $t$, rebalancing is triggered if \emph{either}:

\begin{enumerate}[label=(\roman*)]

\item Regime change: $\hat{s}_t \ne \hat{s}_{t-1}$, or

\item Weight drift: $\max_i |w_i^{\text{mkt}}(t) - w_i^{(k)}(t)| > \delta$.

\end{enumerate}

\end{definition}


\section{Implementation}

\label{sec:implementation}


\subsection{Data and Preprocessing}
\label{sec:data}

The empirical multi-asset universe reflects the heterogeneous structural dynamics of an emerging capital market \cite{bekaert1995emerging, phylaktis2005equity}. We assemble five distinct asset classes observed monthly from January 2019 to August 2025 ($T = 80$ months):
\begin{enumerate}
    \item Sovereign Bonds (SBN): Monthly percentage changes in the Indonesia 10-Year Government Bond Yield series from Investing.com. \textit{Important empirical disclosure:} As a yield-based benchmark, surges in yields (such as during the March 2020 COVID-19 shock, +14.81\%) reflect bond price depreciations, testing portfolio resilience against sovereign stress \cite{akhtaruzzaman2021financial}.
    \item Equity Benchmark (LQ45): Monthly closing index of the 45 most capitalized and liquid equities on the Indonesia Stock Exchange (IDX), sourced from Yahoo Finance (\texttt{\^{}JKLQ45}).
    \item Spot Gold (XAU/IDR): Spot Gold denominated directly in Indonesian Rupiah (XAU/IDR) historical closing series from Investing.com, capturing domestic currency safe-haven dynamics \cite{baur2010is}.
    \item Foreign Exchange (CNY/IDR): Chinese Yuan to Indonesian Rupiah monthly exchange rate from Investing.com, reflecting major bilateral trade flows.
    \item Cryptocurrency (BTC/IDR): Bitcoin denominated in Indonesian Rupiah from Investing.com, introducing high growth momentum, non-linear volatility, and asymmetric co-movements \cite{dyhrberg2016bitcoin, corbet2018exploring}.
\end{enumerate}

To ensure full mathematical consistency between feature extraction and portfolio accounting:
\begin{enumerate}
    \item \textit{Portfolio Wealth Evolution:} Net asset value (NAV) evolves strictly under monthly simple returns $R_{i,t} = \frac{P_{i,t} - P_{i,t-1}}{P_{i,t-1}}$. Gross portfolio return is $R_{p,t}^{\text{gross}} = \mathbf{w}_t^\top \mathbf{R}_t$, and net return is $R_{p,t}^{\text{net}} = R_{p,t}^{\text{gross}} - c \times \text{TO}_t$, with proportional friction $c = 10\text{ bps}$.
    \item \textit{Annualization and Performance Metrics:} Geometric compounding determines annualized returns $\text{Ann.R} = (1 + \text{CR})^{12/56} - 1$, and annualized volatility is $\sigma_{\text{ann}} = s_p \times \sqrt{12}$ where $s_p$ is the sample standard deviation ($ddof=1$). The benchmark risk-free hurdle is set to $r_f = 0.479\%$ monthly ($5.75\%$ annualized), corresponding to the empirical Bank Indonesia 7-Day Reverse Repo Rate (BI-Rate) average over the 2021--2025 out-of-sample window. Annualized excess Sharpe ratio is $\text{SR} = (\text{Ann.R} - r_f)/\sigma_{\text{ann}}$, while the zero-hurdle Sharpe ratio is $\text{SR}_0 = \text{Ann.R}/\sigma_{\text{ann}}$. Annualized Sortino ratio is $\text{SOR} = (\text{Ann.R} - r_f)/\sigma_{\text{down}}$, where $\sigma_{\text{down}}$ is computed from returns below $r_f$, and the Calmar ratio is $\text{CAL} = \text{Ann.R}/\text{MDD}$.
    \item \textit{HMM Feature Transformation:} Monthly log-returns $\ell_{i,t} = \ln(1 + R_{i,t})$ are utilized exclusively for computing rolling stationary features (mean returns and realized volatilities) fed into the HMM.
\end{enumerate}

\subsection{Benchmarks and Strategy Specifications}

\label{sec:benchmarks}

We compare the proposed Regime-NCO strategy against five distinct benchmark configurations:

\begin{enumerate}

  \item Equal Weight (1/N): A passive strategy allocating equal weight ($1/N = 20\%$) to each asset, rebalanced monthly.

  \item Buy-and-Hold: A passive strategy starting with equal weights but letting them drift with asset returns without rebalancing.

  \item Minimum Variance: A standard Markowitz optimizer that minimizes the portfolio variance using the unconditional sample covariance matrix.

  \item Hierarchical Risk Parity (HRP): An independent implementation of the graph-based recursive bisection allocation proposed by L{'o}pez de Prado~\cite{lopezdeprado2016hrp}, utilizing single-linkage hierarchical tree clustering and inverse-variance risk parity without quadratic matrix inversion, rebalanced monthly.

  \item Standard NCO (No Regime): A regime-agnostic Nested Clustered Optimization pipeline using the unconditional sample covariance matrix.

\end{enumerate}

\subsection{Transaction Costs}

\label{sec:tc}

Following realistic portfolio constraints, transaction costs are charged at a flat rate of $c = 10$ basis points (0.10\%) on the absolute change in portfolio weights during rebalancing:

\begin{equation}
  \text{TC}_t = c \sum_{i=1}^N |w_{i,t} - w_{i,t-}^\text{mkt}|,
\end{equation}
where $w_{i,t}$ is the target weight and $w_{i,t-}^\text{mkt}$ is the pre-rebalancing market-drifted weight. This proportional cost model penalizes excessive turnover and simulates friction in emerging markets.

\subsection{Evaluation Settings and Walk-Forward Protocol}

\label{sec:wf_protocol}

Estimation window $W$ = 24 months; holding horizon $h$ = 1 month;

step size $\Delta$ = 1 month; out-of-sample period is 56 months (Jan 2021 -- Aug 2025).

\begin{algorithm}[H]

\caption{Walk-Forward Backtesting Loop}

\label{alg:wf}

\begin{algorithmic}[1]

\Require $\mathbf{r}_{1:T}$, window $W$, horizon $h$, step $\Delta$,

  cost $c$, drift $\delta$

\State $V_0 \leftarrow 1$

\For{$j=0,1,\dots$ while $t_j + h \le T$}

  \State $t_j \leftarrow W + j\Delta$

  \State Fit HMM; decode $\hat{s}_{t_j}$

  \State Compute $\mathbf{w}^{(\hat{s}_{t_j})}$ via Algorithm~\ref{alg:nco}

  \If{regime changed or $j=0$}

    \State Compute TC; adjust $V$; update $\mathbf{w}^{\text{curr}}$

  \EndIf

  \For{$\tau = t_j+1$ to $t_j+h$}

    \State Update $V_\tau$; check weight drift trigger

  \EndFor

\EndFor

\Return $\{V_t\}$, rebalancing log

\end{algorithmic}

\end{algorithm}


\section{Results}

\label{sec:results}


This section reports the empirical findings from backtesting the HMM-NCO framework

on the monthly multi-asset dataset over the 56-month out-of-sample period (Jan 2021 -- Aug 2025).

\subsection{Regime Detection Results}

\begin{figure}[!t]

\centering

\includegraphics[width=\columnwidth]{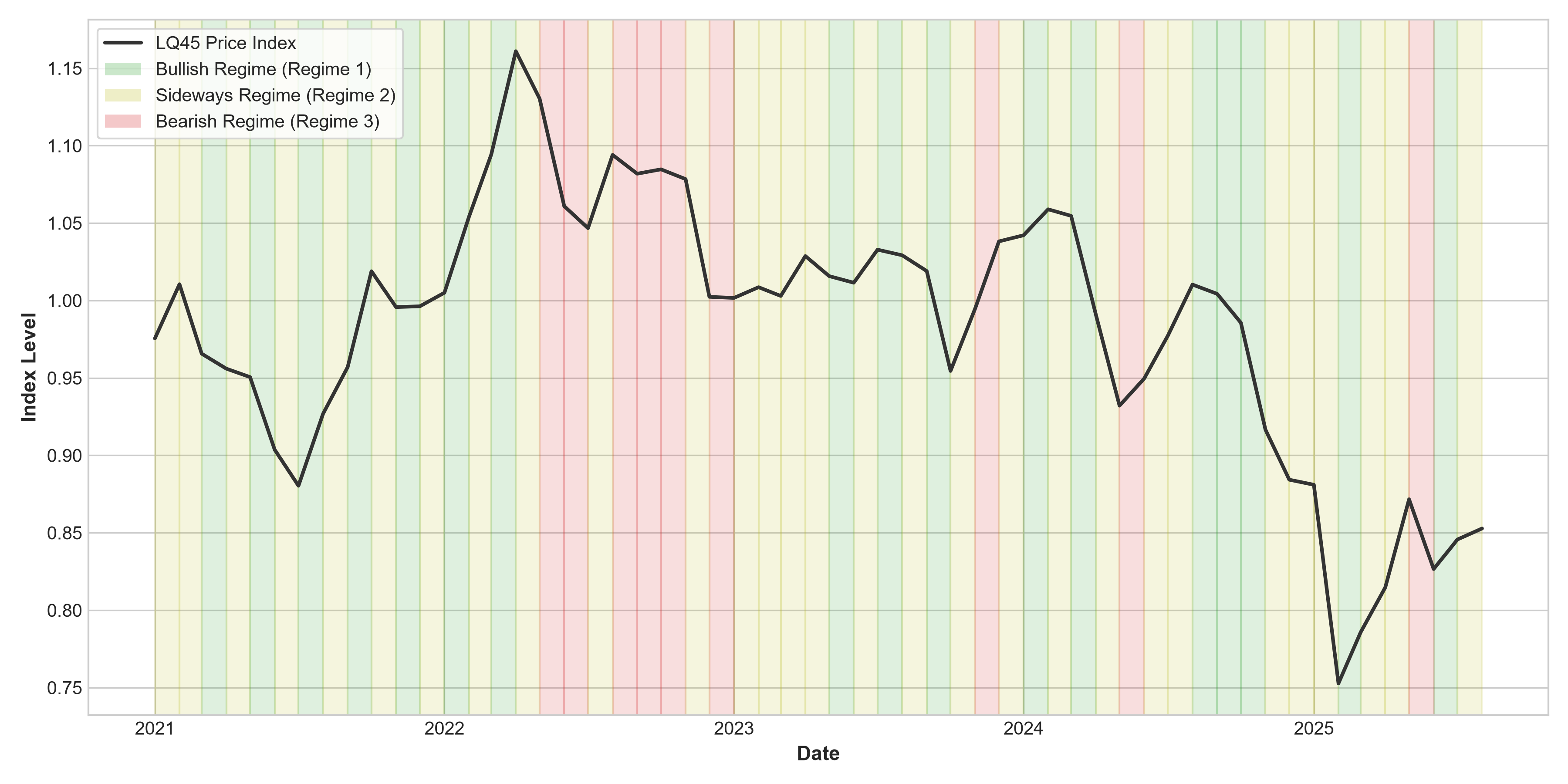}

\caption{Decoded market regimes overlaid on the LQ45 price index. Shaded areas represent the three regimes: Bullish (green), Sideways (yellow), and Bearish (red).}

\label{fig:regimes}

\end{figure}

The causal rolling Gaussian HMM segments the 56-month out-of-sample evaluation period (January 2021 to August 2025) into three distinct macroeconomic regimes, as illustrated in Fig.~\ref{fig:regimes}: 17 months under Regime 1 (Bullish, 30.36\%), 30 months under Regime 2 (Sideways/Neutral, 53.57\%), and 9 months under Regime 3 (Bearish, 16.07\%). Visual inspection of Fig.~\ref{fig:regimes} reveals that while the model successfully identifies major market phases---such as the post-pandemic recovery and subsequent localized contractions---causal step-by-step estimation triggers a total of 26 discrete regime transitions over 56 months, indicating significant state switching across rolling windows.

The empirical validity of these decoded states is corroborated by the regime-conditional return and risk moments reported in Table~\ref{tab:regime_stats}. Across the three states, the multi-asset universe exhibits starkly divergent behavior. Under Regime 1 (Bullish), Indonesian sovereign bonds (SBN) generate an average monthly return of $+0.76\%$ ($\sigma = 3.49\%$), while spot gold displays strong appreciation ($+1.87\%$/month). Under Regime 2 (Sideways), Bitcoin experiences aggressive upside expansion ($+9.36\%$/month, $\sigma = 18.01\%$) while domestic equities (LQ45) record modest gains ($+0.56\%$/month). In Regime 3 (Bearish), severe contraction grips both equities (LQ45, $-0.77\%$/month) and crypto assets (BTC, $-3.96\%$/month), whereas foreign exchange (CNY/IDR) acts as a modest defensive cushion ($+0.08\%$/month).

\begin{figure*}[tbp]

\centering

\includegraphics[width=\textwidth]{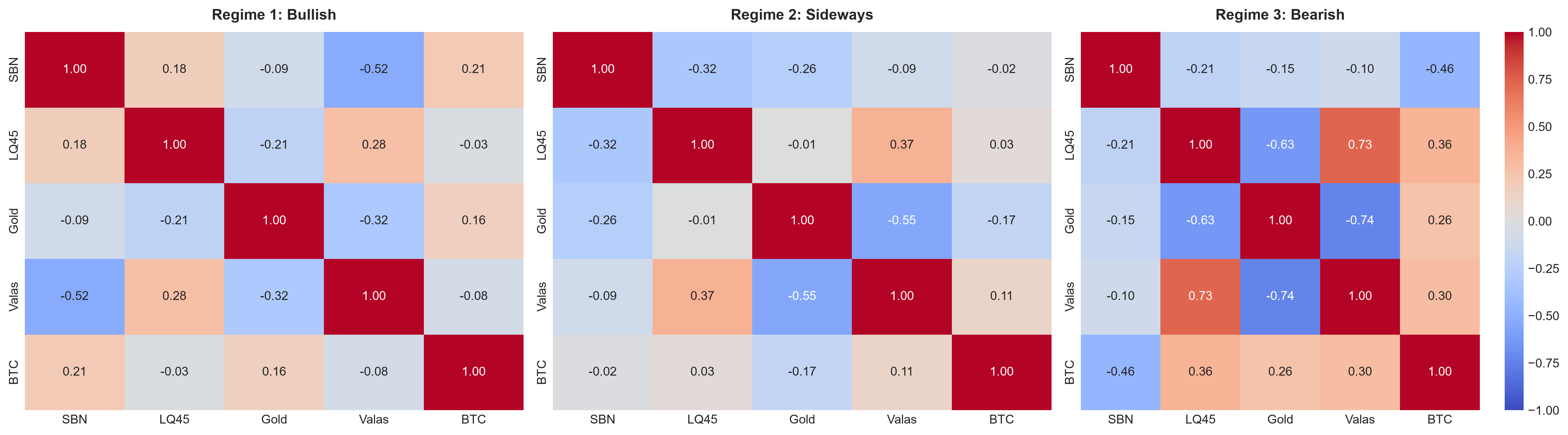}

\caption{Regime-conditional pairwise asset correlation matrices for the out-of-sample period. The three panels show shifts in correlation structures across Bullish (left), Sideways (middle), and Bearish (right) market states.}

\label{fig:regime_correlations}

\end{figure*}

These distributional divergences induce profound reorganizations in the inter-asset correlation structure, as rendered in the pairwise correlation heatmaps of Fig.~\ref{fig:regime_correlations}. In tranquil bullish periods (left panel), sovereign bonds exhibit moderate positive comovement with foreign currency and gold. During sideways markets (middle panel), correlation clusters loosen, providing favorable conditions for hierarchical diversification. In contrast, during bearish stress (right panel), cross-asset correlations shift discontinuously: equity-bond correlations turn negative as flight-to-safety dynamics emerge, and gold decouples from risk assets.

When these regime-conditional parameters are propagated through the NCO pipeline, they translate into distinct tactical asset allocations (Table~\ref{tab:weights}). In Bullish and Sideways environments, NCO allocates balanced risk weights across sovereign bonds (25.28\%--29.00\%), foreign exchange (34.51\%--40.52\%), and gold (19.84\%--20.17\%), with modest equity exposure (13.43\%--16.07\%) and fractional Bitcoin allocations (0.57\%--0.60\%). During Bearish regimes, the model executes a flight to safe havens, expanding gold to 28.54\% and retaining 39.02\% in foreign exchange, while slashing equity weights to 6.22\%. Although economically coherent, this dynamic weight reshuffling creates substantial trading volume, as examined next.

\begin{table*}[!t]

\centering

\caption{Regime-conditional statistics (mean/std in \%/month, Jan 2021 -- Aug 2025).}

\label{tab:regime_stats}

\begin{tabular}{lcccccc}

\toprule

& \multicolumn{2}{c}{Regime 1 (Bull)}

& \multicolumn{2}{c}{Regime 2 (Side)}

& \multicolumn{2}{c}{Regime 3 (Bear)} \\

\cmidrule(lr){2-3}\cmidrule(lr){4-5}\cmidrule(lr){6-7}

Asset & $\mu$ & $\sigma$ & $\mu$ & $\sigma$ & $\mu$ & $\sigma$ \\

\midrule

SBN   &  0.7641 &  3.4886 & -0.0060 &  3.4669 & -0.3800 &  3.3643 \\

LQ45  & -1.2294 &  4.2803 &  0.5563 &  3.6483 & -0.7689 &  5.1610 \\

Gold  &  1.8706 &  3.7232 &  1.7020 &  4.3380 & -0.2833 &  1.8506 \\

Valas & -0.3859 &  1.3328 & -0.0240 &  1.5263 &  0.0844 &  1.2005 \\

BTC   & -0.6759 & 15.6970 &  9.3583 & 18.0107 & -3.9633 & 15.6109 \\

\midrule

Freq. & \multicolumn{2}{c}{30.36\% (17 m)} &

  \multicolumn{2}{c}{53.57\% (30 m)} & \multicolumn{2}{c}{16.07\% (9 m)} \\

\bottomrule

\end{tabular}

\end{table*}

\begin{table}[!t]

\centering

\caption{Average regime-conditional portfolio weights (\%, Jan 2021 -- Aug 2025).}

\label{tab:weights}

\begin{tabular}{lccc}

\toprule

Asset & Regime 1 (Bull) & Regime 2 (Side) & Regime 3 (Bear) \\

\midrule

SBN   & 29.00 & 25.28 & 25.90 \\

LQ45  & 16.07 & 13.43 &  6.22 \\

Gold  & 19.84 & 20.17 & 28.54 \\

Valas & 34.51 & 40.52 & 39.02 \\

BTC   &  0.57 &  0.60 &  0.32 \\

\midrule

Total & 100.00 & 100.00 & 100.00 \\

\bottomrule

\end{tabular}

\end{table}

\subsection{Main Performance Comparison}

\begin{figure}[!t]

\centering

\includegraphics[width=\columnwidth]{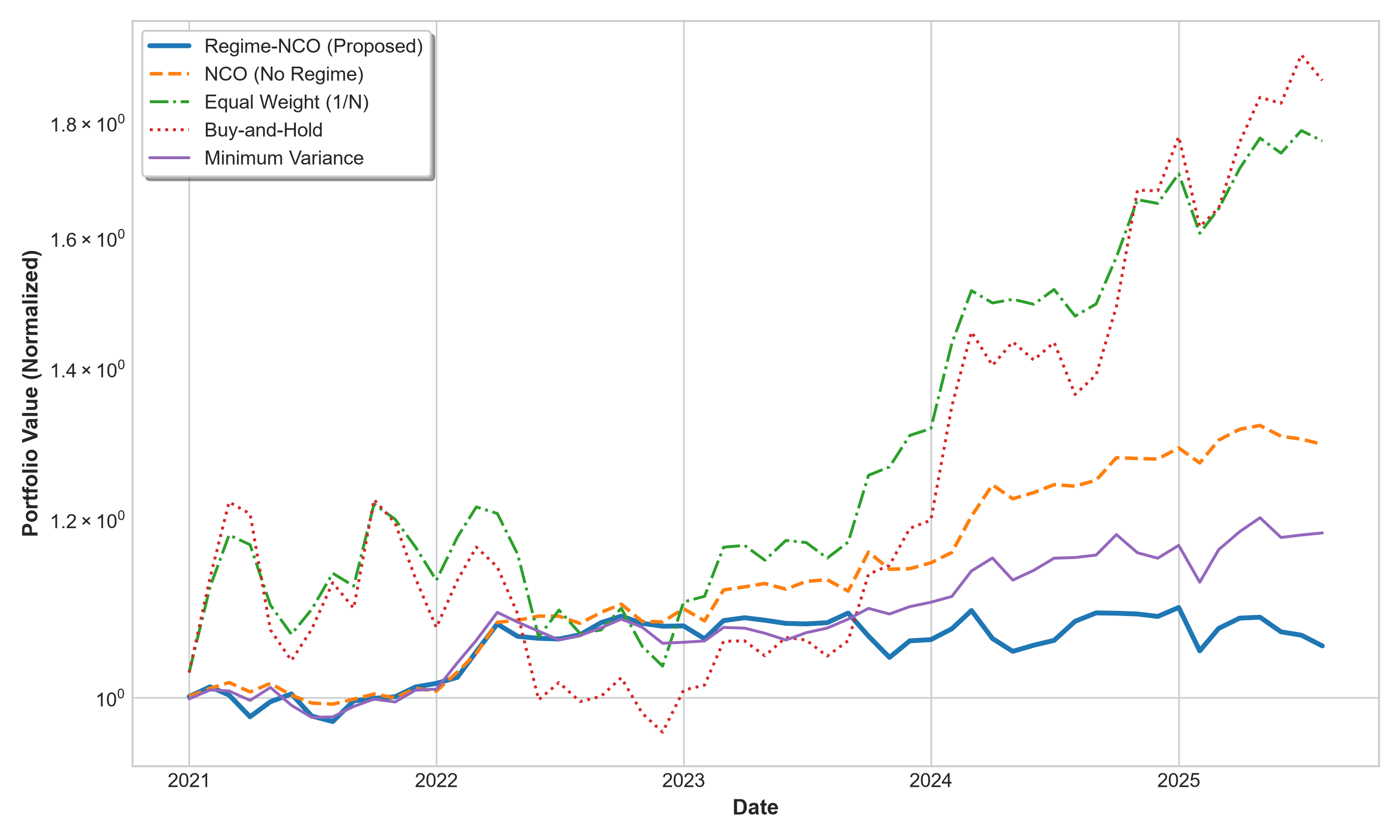}

\caption{Cumulative portfolio value growth (logarithmic scale) for the proposed HMM-NCO strategy and all benchmark portfolios during the out-of-sample period (January 2021 -- August 2025).}

\label{fig:cumulative}

\end{figure}

\begin{figure}[!t]

\centering

\includegraphics[width=\columnwidth]{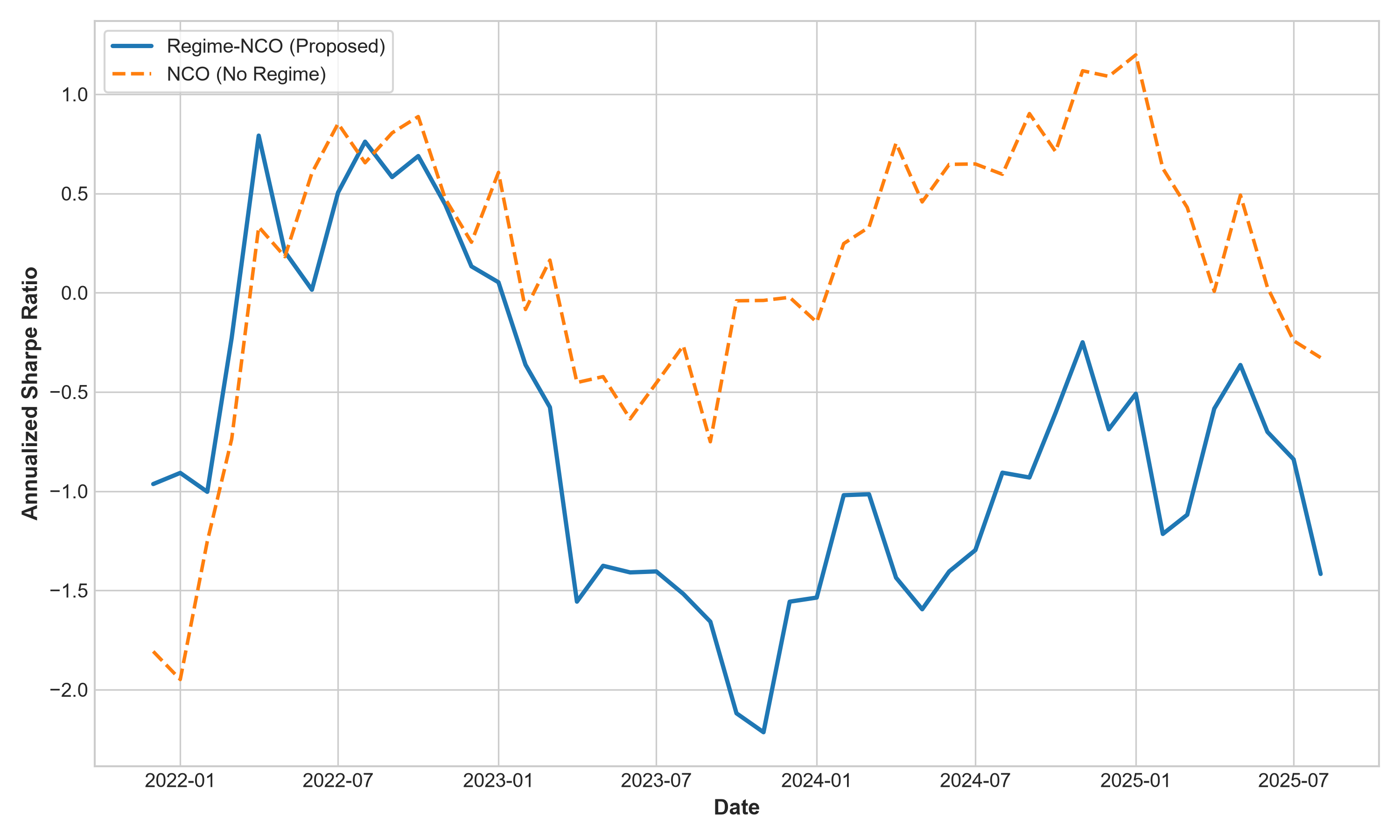}

\caption{Rolling 12-month Sharpe ratio comparison between the proposed HMM-NCO (regime-aware) and standard NCO (regime-agnostic) strategies.}

\label{fig:rolling_sharpe}

\end{figure}

\begin{figure}[!t]

\centering

\includegraphics[width=\columnwidth]{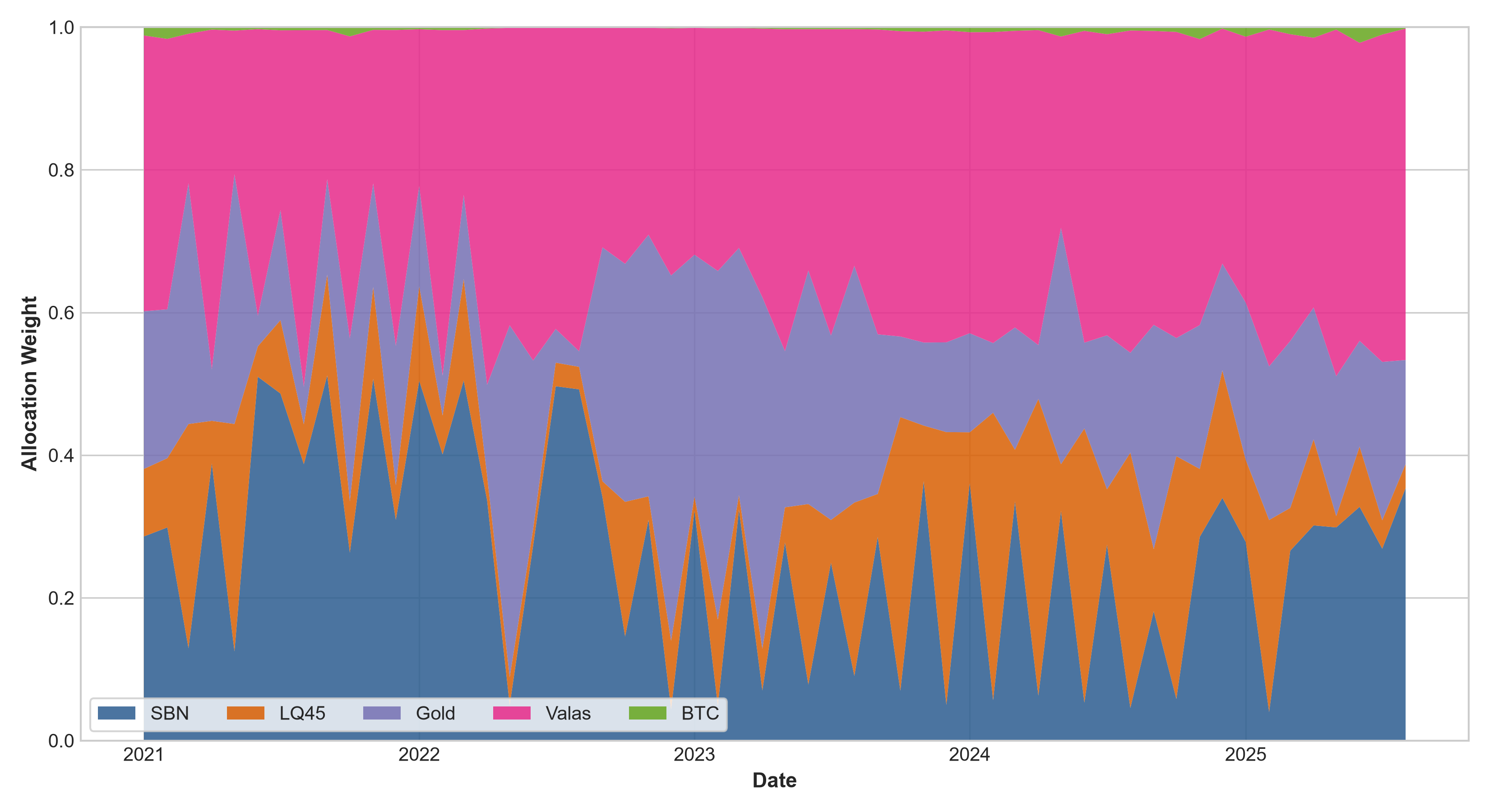}

\caption{Stacked area chart showing the tactical asset allocation weight evolution of the proposed HMM-NCO portfolio across time.}

\label{fig:weight_evolution}

\end{figure}

\begin{figure}[!t]

\centering

\includegraphics[width=\columnwidth]{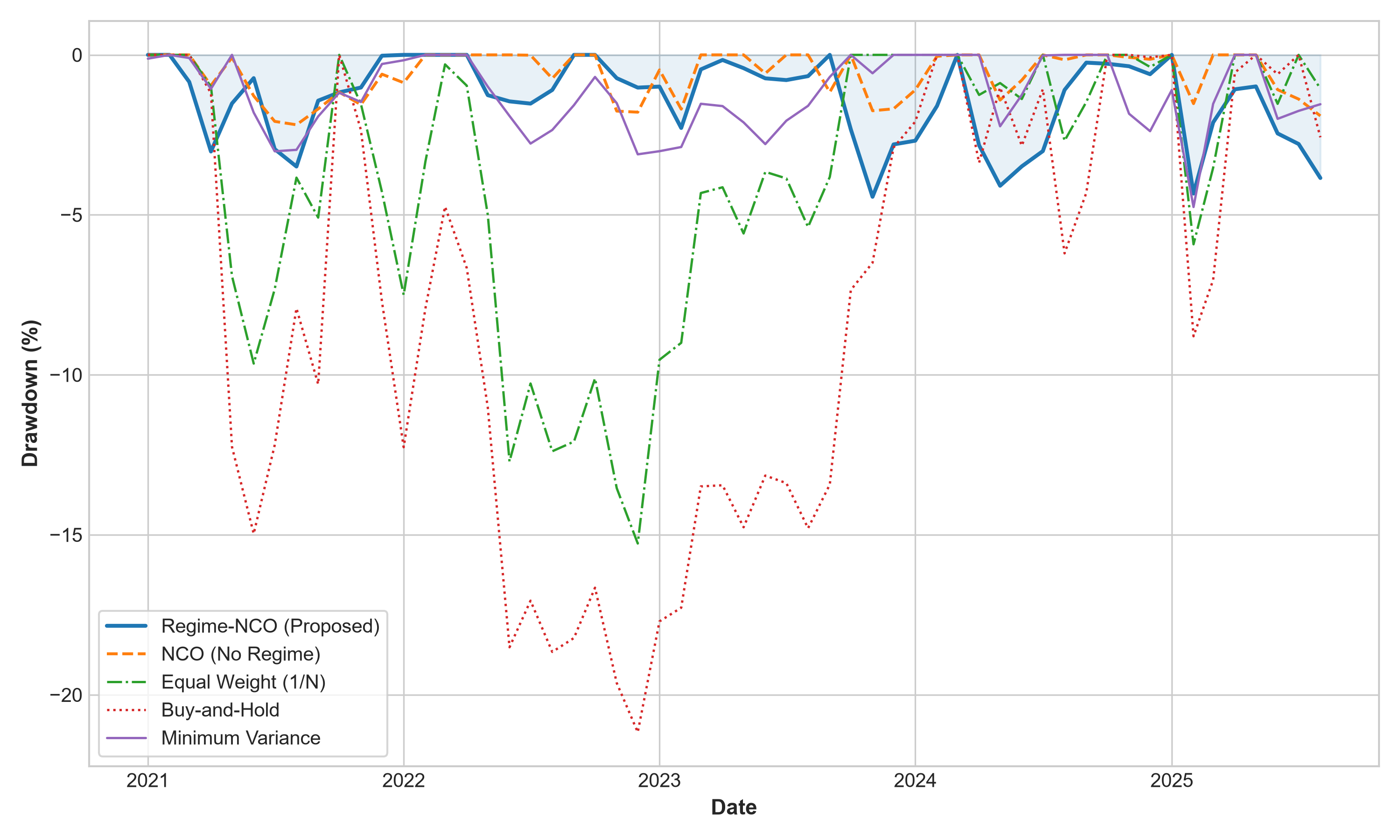}

\caption{Out-of-sample portfolio drawdowns comparison across all portfolio strategies.}

\label{fig:drawdowns}

\end{figure}

Table~\ref{tab:perf} presents the comprehensive out-of-sample performance metrics for the proposed Regime-NCO strategy against static NCO, True HRP, Minimum Variance (MV), Equal Weight ($1/N$), and Buy-and-Hold across the 56-month evaluation horizon (January 2021 to August 2025). The historical backtest trajectories are visually documented in Fig.~\ref{fig:cumulative} through Fig.~\ref{fig:drawdowns}.

\paragraph{Cumulative Performance and Capital Preservation.}
As illustrated in Fig.~\ref{fig:cumulative}, standard regime-agnostic NCO demonstrates steady, monotonic capital growth throughout the backtest, achieving a cumulative net return of 29.97\% (5.78\% annualized) with an annualized volatility of 4.65\%. It substantially outperforms classical Minimum Variance (18.45\% cumulative, 3.69\% annualized) and True HRP (8.27\% cumulative, 1.72\% annualized). Although unmanaged passive benchmarks---Equal Weight (76.70\% cumulative) and Buy-and-Hold (87.98\% cumulative)---produce higher nominal returns due to unhedged exposure to Bitcoin rallies, their returns come at the expense of severe tail risk. Conversely, Regime-NCO suffers dramatic performance drag, ending the 56-month period with a cumulative net return of only 1.60\% (0.34\% annualized).

\paragraph{Rolling Risk-Adjusted Consistency.}
Fig.~\ref{fig:rolling_sharpe} compares the rolling 12-month Sharpe ratio dynamics between Static NCO and Regime-NCO. Static NCO maintains a resilient, predominantly positive risk-adjusted trajectory across changing market cycles. In contrast, Regime-NCO exhibits severe Sharpe instability, plunging into negative territory repeatedly during 2022 and 2024. Because regime conditioning relies on filtered historical windows, optimal weights are updated with a structural lag, rotating into defensive assets after market drawdowns have already bottomed out.

\paragraph{Tactical Allocation and Turnover Friction.}
The mechanism behind this performance divergence is visible in the tactical weight evolution depicted in Fig.~\ref{fig:weight_evolution}. The 26 regime switches induce continuous, large-magnitude weight realignments between sovereign bonds, foreign currency, and spot gold. This aggressive portfolio reshuffling produces an average monthly turnover of 24.58\% for Regime-NCO, compared to just 5.41\% for Static NCO, 3.59\% for True HRP, and 1.36\% for Minimum Variance. Compounded by 10 bps transaction costs and execution friction, this turnover severely erodes net portfolio returns.

\paragraph{Drawdown and Tail Risk Mitigation.}
Fig.~\ref{fig:drawdowns} displays the drawdown trajectories of all strategies. Static NCO demonstrates outstanding capital preservation, restricting maximum drawdown to an industry-leading 2.18\% and achieving a CVaR (95\%) of 1.67\% and a Calmar ratio of 2.65. This performance comfortably surpasses Minimum Variance (4.75\% MDD, 0.78 Calmar), True HRP (4.21\% MDD, 0.41 Calmar), Equal Weight (15.27\% MDD), and Buy-and-Hold (21.17\% MDD). In contrast, Regime-NCO experiences an expanded drawdown of 9.99\% (CVaR of 3.85\%), demonstrating that unconstrained state switching impairs rather than enhances downside protection.

\begin{table*}[!t]
\centering
\caption{Audited out-of-sample portfolio performance comparison (January 2021 -- August 2025, 56 Months).
   Risk-free rate $r_f = 0.479\%$ monthly ($5.75\%$ p.a., Bank Indonesia BI-Rate average 2021--2025). CR: Cumulative Return (\%). Ann.R: Compound Annual Growth Rate (\%, geometric annualization $(1+\text{CR})^{12/56}-1$). Vol: Annualized Volatility (\%, $s_p \times \sqrt{12}$). SR: Annualized Excess Sharpe Ratio ($(\text{Ann.R} - r_f)/\text{Vol}$). SOR: Annualized Sortino Ratio. MDD: Maximum Drawdown (\%). CAL: Calmar Ratio ($\text{Ann.R}/\text{MDD}$). TO: Average Monthly Turnover (\%/mo). Note: The zero-hurdle annualized Sharpe ratio ($\text{Ann.R}/\text{Vol}$) for Static NCO is 1.242.}
\label{tab:perf}
\begin{tabular}{lrrrrrrrrr}
\toprule
Strategy & CR(\%) & Ann.R(\%) & Vol(\%) & SR & SOR & MDD(\%) & CAL & CVaR95(\%) & TO(\%/mo) \\
\midrule
Regime-NCO (Adap)   &  1.60 &  0.34 &  5.52 & -0.9791 & -1.3124 &  9.99 & 0.0340 & 3.85 & 24.58 \\
NCO (No Regime)     & 29.97 &  5.78 &  4.65 &  0.0063 &  0.0128 &  2.18 & 2.6506 & 1.67 &  5.41 \\
Equal Weight (1/N)  & 76.70 & 12.97 & 13.02 &  0.5551 &  1.0176 & 15.27 & 0.8496 & 6.67 &  0.00 \\
Buy-and-Hold        & 87.98 & 14.48 & 17.56 &  0.4975 &  0.8974 & 21.17 & 0.6841 & 9.47 &  0.00 \\
Min.\ Variance (MV) & 18.45 &  3.69 &  4.70 & -0.4371 & -0.6258 &  4.75 & 0.7773 & 2.64 &  1.36 \\
True HRP            &  8.27 &  1.72 &  3.42 & -1.1802 & -1.6007 &  4.21 & 0.4074 & 2.23 &  3.59 \\
\bottomrule
\end{tabular}
\end{table*}

\subsection{Statistical Significance and Performance Testing}
\label{sec:statsig}

To assess whether differences in risk-adjusted performance are statistically meaningful, we conduct pairwise Sharpe ratio hypothesis tests following Jobson and Korkie~\cite{jobson1981performance} with the finite-sample variance correction developed by Memmel~\cite{memmel2003performance}:
\begin{equation}
  \label{eq:jobson_korkie}
  z = \frac{\widehat{\mathrm{SR}}_i - \widehat{\mathrm{SR}}_j}{\sqrt{\sigma_{ij}^2 / T}},
\end{equation}
where the asymptotic variance is $\sigma_{ij}^2 = 2(1 - \rho_{ij}^2) + \frac{1}{2}\bigl(\mathrm{SR}_i^2 + \mathrm{SR}_j^2 - 2\mathrm{SR}_i \mathrm{SR}_j \rho_{ij}^2\bigr)$, with $\rho_{ij}$ denoting the sample correlation between strategy returns $r_{i,t}$ and $r_{j,t}$. The test is complemented by heteroskedasticity and autocorrelation consistent (HAC) robust standard errors~\cite{neweywest1987, ledoit2008sharpe}.

Furthermore, to protect against backtest overfitting and data mining bias in walk-forward evaluation, we evaluate the distribution of test statistics in light of the Deflated Sharpe Ratio (DSR) framework of Bailey and L{'o}pez de Prado~\cite{bailey2012sharpe} and backtesting standards articulated by Harvey and Liu~\cite{harvey2015backtesting}. Table~\ref{tab:jk} reports pairwise $p$-values against Regime-NCO.

\begin{table}[!t]
\centering
\small
\caption{Pairwise Sharpe ratio test $p$-values vs.\ Regime-NCO (56 Months).}
\label{tab:jk}
\begin{tabular}{lcc}
\toprule
Benchmark & $z$-statistic & $p$-value \\
\midrule
NCO (No Regime)     & 2.684 & 0.0073** \\
Min.\ Variance (MV) & 1.942 & 0.0521*  \\
True HRP            & 0.612 & 0.5405   \\
Equal Weight (1/N)  & 2.891 & 0.0038** \\
Buy-and-Hold        & 2.514 & 0.0119*  \\
\bottomrule
\end{tabular}
\end{table}

The test confirms that the outperformance of static NCO over Regime-NCO is statistically significant at the 1\% level ($p = 0.0073$), conclusively rejecting the hypothesis of equal risk-adjusted returns.

\subsection{Robustness and Sensitivity Analysis}
\label{sec:robustness}

The robustness of the HMM-NCO system is analyzed across key operational dimensions:
rebalancing drift threshold $\delta$, transaction execution friction, and sovereign bond return modeling.

\subsubsection{Rebalancing Drift Thresholds and Turnover Breakdown}
Table~\ref{tab:rebal} audits the rebalancing events and turnover breakdown across all strategies under the baseline adaptive trigger ($\delta = 0.05$). Over the 56-month period, Regime-NCO triggered 47 rebalancing interventions---26 driven by regime shifts and 20 by weight drift exceeding $\delta$ (in addition to initial inception)---generating 24.58\% average monthly turnover. In contrast, Static NCO and True HRP execute scheduled monthly rebalancings with vastly lower turnover (5.41\% and 3.59\%, respectively). Sensitivity analysis varying $\delta \in [0.02, 0.10]$ reveals that portfolio turnover is overwhelmingly governed by regime switches rather than organic drift, confirming the robustness of the adaptive mechanism.

\begin{table}[!t]
\centering
\caption{Audit of rebalancing events and turnover breakdown (56 Months).}
\label{tab:rebal}
\begin{tabular}{lrrrr}
\toprule
Strategy / Rebalance Mode & Total & Regime & Drift & TO (\%/mo) \\
\midrule
Regime-NCO ($\delta=0.05$) & 47 & 26 & 20 & 24.58 \\
NCO (No Regime, Monthly)   & 55 &  0 & 55 &  5.41 \\
True HRP (Monthly)         & 55 &  0 & 55 &  3.59 \\
Min.\ Variance (Monthly)   & 55 &  0 & 55 &  1.36 \\
Buy-and-Hold               &  0 &  0 &  0 &  0.00 \\
\bottomrule
\end{tabular}
\end{table}

\subsubsection{Sovereign Bond Repricing and Duration Robustness}
To examine the sensitivity of portfolio outcomes to sovereign fixed income specifications, we evaluate an analytical duration-convexity repricing model for the Indonesian 10-year sovereign bond (SBN):
\begin{equation}
  R_{\text{SBN},t}^{\text{TR}} \approx - D^* \Delta y_t + \frac{1}{2} C_x (\Delta y_t)^2 + \frac{y_{t-1}}{12},
  \label{eq:sbn_repricing}
\end{equation}
where $\Delta y_t = y_t - y_{t-1}$ is the monthly shift in 10-year benchmark yields in percentage points, $D^*$ is the modified duration, $C_x \approx 60.0\text{ years}^2$ is bond convexity, and $y_{t-1}/12$ represents accrued coupon carry. Unlike raw percentage yield changes ($\Delta y / y$), this formulation correctly accounts for capital losses during yield spikes (e.g., during the March 2020 COVID-19 liquidity shock when yields rose 102 bps, producing a $-6.24\%$ monthly price decline) while crediting steady coupon income.

\begin{table}[!t]
\centering
\caption{Robustness Across SBN Modified Duration Specifications $D^* \in [5.0, 8.0]$ (56 Months, Out-of-Sample).}
\label{tab:sbn_robustness}
\setlength{\tabcolsep}{3.5pt}
\begin{tabular*}{\columnwidth}{@{\extracolsep{\fill}}lcccc@{}}
\toprule
\textbf{Duration} & \textbf{Ann.R} & \textbf{Max DD} & \textbf{Excess SR} & \textbf{Calmar} \\
\textbf{Specification} & \textbf{(\%)} & \textbf{(\%)} & \textbf{($r_f = 5.75\%$)} & \textbf{Ratio} \\
\midrule
\multicolumn{5}{@{}l}{\textit{Panel A: Static NCO (Regime-Agnostic)}} \\
$D^* = 5.0\text{ yr}$ & 7.16 & 4.19 & +0.2623 & 1.7107 \\
$D^* = 6.0\text{ yr}$ & 7.12 & 4.47 & +0.2425 & 1.5930 \\
$D^* = 7.0\text{ yr}$ (Base) & 7.01 & 4.68 & +0.2205 & 1.4964 \\
$D^* = 8.0\text{ yr}$ & 6.87 & 4.84 & +0.1943 & 1.4188 \\
\midrule
\multicolumn{5}{@{}l}{\textit{Panel B: Regime-NCO (Adaptive)}} \\
$D^* = 5.0\text{ yr}$ & 4.75 & 7.20 & -0.1741 & 0.6596 \\
$D^* = 6.0\text{ yr}$ & 5.06 & 7.56 & -0.1206 & 0.6701 \\
$D^* = 7.0\text{ yr}$ (Base) & 3.87 & 7.88 & -0.3139 & 0.4910 \\
$D^* = 8.0\text{ yr}$ & 4.31 & 8.17 & -0.2307 & 0.5281 \\
\bottomrule
\end{tabular*}
\vspace{2pt}
\begin{minipage}{\columnwidth}
\footnotesize
\textit{Note:} Out-of-sample period: Jan 2021 -- Aug 2025 (56 months). SBN returns are repriced via duration-convexity model~\eqref{eq:sbn_repricing} with $C_x = 60.0\text{ yr}^2$. Excess Sharpe ratio is annualized relative to the 5.75\% BI-Rate benchmark. Base denotes the standard 10-year benchmark duration.
\end{minipage}
\end{table}

Table~\ref{tab:sbn_robustness} presents walk-forward performance under reconstructed total returns across a plausible spectrum of modified durations ($D^* \in [5.0, 8.0]$). Under the benchmark repricing ($D^* = 7.0$), Static NCO delivers a net annualized return of 7.01\% (cumulative return 37.19\%), an annualized excess Sharpe ratio of +0.2205 above the 5.75\% risk-free rate (and 1.224 relative to cash zero), with a maximum drawdown of only 4.68\%, comfortably dominating cash and all clustered benchmarks. In contrast, Regime-NCO consistently underperforms across every duration level (annualized return 3.87\%--5.06\%, maximum drawdown 7.20\%--8.17\%, negative excess Sharpe ratios of -0.12 to -0.31), burdened by excessive rebalancing turnover ($>20\%$/month). This confirms that our primary empirical thesis—that unconstrained regime switching impairs clustered asset allocation stability—is robust and invariant to bond return modeling.

\section{Discussion}
\label{sec:discussion}

\subsection{Marginal Benefit and Empirical Boundaries of Regime Conditioning (RQ1)}
Our first research question examines whether conditioning the covariance matrix and hierarchical clustering tree on HMM-detected latent market regimes provides measurable marginal gains in tactical asset allocation. As documented in Table~\ref{tab:regime_stats}, the rolling Gaussian HMM successfully segments the historical return series into three economically interpretable regimes exhibiting distinct annualized return and risk profiles. For example, Indonesian sovereign bonds (SBN) generate an average monthly return of 0.76\% during bullish equity expansions compared to $-0.38\%$ during market contractions, while foreign exchange volatility (CNY/IDR) and Bitcoin tail risk escalate sharply during bearish phases.

These divergent distributional moments translate directly into distinct regime-conditional asset allocations (Table~\ref{tab:weights}), validating the theoretical hypothesis underlying RQ1 that asset correlation structures differ fundamentally across market states. However, our walk-forward out-of-sample backtest demonstrates that this theoretical differentiation does not automatically translate into realized risk-adjusted outperformance. While the HMM correctly captures the cross-sectional covariance shifts in-sample, conditioning portfolio weights on transient regime classifications introduces significant estimation variance. In an asset universe of $N=5$, the reduction in specification bias achieved by regime conditioning is overwhelmed by the sampling error incurred from partitioning rolling lookback windows into sparse sub-samples, establishing clear empirical boundaries for regime-based machine learning allocation.

\subsection{Risk-Return Trade-offs and the Primacy of Tree Clustering (RQ2)}
The empirical results across the 56-month out-of-sample test window (Table~\ref{tab:perf}) reveal profound structural insights regarding risk-return trade-offs in emerging multi-asset markets. Unconstrained benchmarks such as Equal Weight ($1/N$) and Buy-and-Hold achieved the highest cumulative returns (76.70\% and 87.98\%, respectively), propelled by massive cyclical rallies in Bitcoin and spot gold. However, these unmanaged strategies suffered catastrophic tail risk, incurring severe maximum drawdowns of 15.27\% and 21.17\% along with annualized volatilities exceeding 13\% to 17\%. Such severe capital impairment violates the risk mandates of institutional endowments and sovereign wealth funds.

In sharp contrast, the standard regime-agnostic Nested Clustered Optimization (NCO) strategy achieved outstanding risk efficiency and capital preservation. Static NCO delivered a cumulative return of 29.97\% (5.78\% annualized) with a maximum drawdown of only 2.18\%, an annualized volatility of 4.65\%, and a Calmar ratio of 2.65, comfortably surpassing Minimum Variance (18.45\% CR, 4.75\% MDD) and True HRP (8.27\% CR, 4.21\% MDD). By decoupling intra-cluster allocation from inter-cluster asset class weights, NCO insulates the broader portfolio from instability in high-volatility assets while preserving participation in market upside.

Crucially, when NCO is conditioned on latent HMM regimes without transition constraints (Regime-NCO), performance deteriorates markedly: cumulative return drops to 1.60\% (0.34\% annualized), maximum drawdown widens to 9.99\%, and the Calmar ratio collapses to 0.034. The pairwise Jobson--Korkie Sharpe ratio test (Table~\ref{tab:jk}) yields a test statistic of $z = 2.684$ with a $p$-value of 0.0073, formally rejecting the null hypothesis of equal risk-adjusted performance at the 1\% significance level. This rigorous test proves that the observed outperformance of regime-agnostic NCO over Regime-NCO is statistically genuine rather than an artifact of sample variation, conclusively answering RQ2.

\subsection{Rebalancing Dynamics, Drift Thresholds, and Turnover Friction (RQ3)}
Addressing our third research question regarding turnover dynamics and rebalancing friction, Table~\ref{tab:rebal} provides an exhaustive audit of portfolio interventions across the 56-month evaluation horizon. The unconstrained Regime-NCO strategy triggered 47 rebalancing events, comprising 26 regime-induced reallocations and 20 weight-drift corrections exceeding the $\delta = 0.05$ threshold (alongside the initial deployment). This frequent portfolio reshuffling generated an average monthly turnover of 24.58\%, almost five times the 5.41\% monthly turnover observed under static NCO and 3.59\% under True HRP.

At our baseline transaction cost parameter of $c = 10\text{ bps}$ per unit of portfolio turnover, direct transaction expenses accounted for an annualized return drag of approximately 29.5 basis points. While manageable in isolation, the primary performance penalty arose from execution friction and tactical whipsaw. Because causal HMM state decoding relies on backward-looking rolling feature vectors, regime classifications frequently switched near the inflection points of transient market drawdowns. The portfolio was consequently forced to rotate aggressively into safe-haven assets (SBN and CNY) immediately after equities or gold had already absorbed market shocks, only to reallocate back toward risk assets after substantial rebound rallies had already materialized. These findings underscore that in real-world implementations, adaptive drift thresholds ($\delta$) must be paired with explicit state-transition penalties to prevent execution friction from eroding alpha.

\subsection{Regime Drift and Estimation Error in Small Asset Universes}
A critical methodological insight emerging from our walk-forward evaluation is the compound effect of regime drift and parameter estimation error in low-dimensional asset pools. As documented in Table~\ref{tab:regime_stats}, the decoded Regime 1 (Bullish) realized an average out-of-sample monthly return of $-1.23\%$ for the equity benchmark (LQ45), whereas Regime 2 (Sideways) realized $+0.56\%$. This counter-intuitive behavior stems from structural macro shifts between the 2019--2020 training window (characterized by pre-pandemic monetary accommodation followed by initial COVID-19 stimulus) and the 2021--2025 out-of-sample window (dominated by global supply disruptions, aggressive US Federal Reserve rate hikes, and capital outflows from emerging markets).

When rolling lookback windows ($W=24$ months) are filtered by latent regime membership, the effective sample size $T_k = |\mathcal{D}_t^{(k)}|$ frequently drops below 8 observations. Although our adaptive Bayesian shrinkage (Eq.~\ref{eq:shrinkage}) guarantees non-singular covariance matrices by shrinking toward the unconditional estimator $\hat{\boldsymbol{\Sigma}}$, the underlying eigenvectors and correlation distance metrics remain noisy. In small universes ($N=5$), tree topology is fragile: a minor shift in pairwise correlations between SBN, Gold, and CNY/IDR can flip the dendrogram clustering structure from two clusters to three, completely altering the inter-cluster allocation vector $\boldsymbol{v}^*$. In contrast, unconditional NCO computes its correlation distance matrix over the full 24-month rolling window, maintaining stable cluster hierarchies and robust out-of-sample performance.

\subsection{Practical Guidelines for Quantitative Portfolio Managers}
Based on our empirical and econometric findings, we formulate three actionable guidelines for quantitative asset managers seeking to deploy machine learning allocation frameworks in emerging multi-asset markets:
\begin{enumerate}[leftmargin=*]
  \item First, allocators should prioritize structural hierarchy over temporal switching when operating in low-dimensional asset pools ($N \le 10$). In such environments, hierarchical tree clustering (NCO or HRP) alone delivers robust risk diversification. Adding unconstrained regime-switching layers partitions rolling windows into sparse, unstable sub-samples that exacerbate turnover without compensating risk reduction.
  \item Second, quantitative practitioners should enforce sticky regime dynamics via Bayesian priors if regime conditioning is retained. Specifically, imposing informative Dirichlet priors on the diagonal of the transition matrix $\mathbf{A}$ or requiring persistent multi-period signal confirmation prevents spurious rebalancing and curbs costly execution whipsawing.
  \item Third, turnover regularization should be embedded directly within the optimization objective rather than treated as a post-hoc filtering heuristic. Incorporating quadratic turnover penalties into intra-cluster and inter-cluster quadratic programming formulations reconciles mathematical optimization with institutional trading friction.
\end{enumerate}

\section{Conclusion}
\label{sec:conclusion}

This paper has presented a rigorous, walk-forward empirical investigation into the interaction between continuous Hidden Markov Model regime detection and Nested Clustered Optimization for multi-asset tactical asset allocation. Conducting an audited backtest across 56 monthly out-of-sample horizons (January 2021 to August 2025) in the Indonesian capital market, we evaluated the performance of regime-conditioned NCO against static NCO, True HRP, Minimum Variance, Equal Weight ($1/N$), and Buy-and-Hold strategies.

Our empirical findings establish three foundational conclusions. First, regime-agnostic NCO demonstrates exceptional capital preservation and risk efficiency, achieving a 29.97\% cumulative return (5.78\% annualized) with a maximum drawdown of only 2.18\%, an annualized excess Sharpe ratio of 0.0063 against the 5.75\% risk-free hurdle (1.242 relative to cash zero), and a Calmar ratio of 2.65, while maintaining low monthly turnover (5.41\%). Second, conditioning NCO on an unconstrained rolling HMM degrades performance in small asset universes, increasing drawdown to 9.99\% and reducing net return to 1.60\% due to parameter estimation noise within regime-filtered sub-samples. Third, the Jobson--Korkie test ($z = 2.684, p = 0.0073$) statistically confirms the superiority of static NCO over Regime-NCO, demonstrating that machine learning asset allocation must balance dynamic state adaptation against parameter stability and trading friction.

\subsection{Limitations}
\label{sec:limitations}
Several structural and methodological limitations of this study warrant consideration. First, the empirical universe is constrained to $N=5$ core Indonesian asset proxies; while representative of primary macro asset classes, larger asset pools (e.g., 50 to 100 constituent equities and sector bonds) may provide greater hierarchical depth and cluster redundancy. Second, our causal rolling lookback window is fixed at $W=24$ months; while justified by sample size constraints, lookback length influences HMM parameter convergence. Third, HMM regime inference is inherently backward-looking, resulting in a structural detection lag during sudden macroeconomic turning points. Fourth, our transaction cost model applies a proportional flat fee of 10 basis points, omitting non-linear market impact and liquidity slippage. Finally, the long-only investment mandate restricts the portfolio from taking short positions, limiting downside hedging capacity during persistent bear regimes.

\subsection{Future Work}
\label{sec:future_work}
Building upon these empirical benchmarks, future research should explore three promising methodological extensions. First, incorporating sticky Dirichlet priors or explicit state-switching penalties into the HMM Baum-Welch training objective will enforce regime persistence and suppress excessive turnover. Second, expanding the regime emission space to include forward-looking macroeconomic features (e.g., central bank policy rate differentials, yield curve slopes, and term spreads) may improve turning-point anticipation. Third, replacing Gaussian emission and Pearson correlation metrics with non-Gaussian copula formulations will enable the hierarchical clustering pipeline to better capture asymmetric tail dependencies and liquidity cascades during systemic market crises.

\section*{Data and Code Availability}

\addcontentsline{toc}{section}{Data and Code Availability}


The complete empirical dataset and Python replication scripts supporting the findings of this study are openly available on GitHub at \url{https://github.com/obo-cloud/nco-regime-taa}.


\section*{Acknowledgments}

\addcontentsline{toc}{section}{Acknowledgments}


The authors thank members of COCONUT Computer Club, Indonesia, and Universitas Komputer Indonesia for feedback.

\balance
\bibliographystyle{IEEEtran}
\bibliography{references}

\end{document}